\documentclass[aps,prb,shownopacs,onecolumn,superscriptaddress]{revtex4}

\usepackage{bm,amsmath,amssymb,mathrsfs,latexsym,graphicx,psfrag,accents,float}
\pdfoutput=1

\usepackage{caption}
\usepackage{indentfirst}
\usepackage{epsfig}
\usepackage[pdftex]{color}
\usepackage{color}
\usepackage[tight]{subfigure}
\definecolor{dred}{rgb}{0.7,0.0,0.0}
\usepackage{dcolumn}% Align table columns on decimal point
\usepackage{bm,bbm}% bold math
\allowdisplaybreaks 
\usepackage{ulem} % to strike things out
\usepackage{hyperref}
\hypersetup{colorlinks=true, linkcolor=blue, citecolor=blue, filecolor=blue, urlcolor=blue, pdftitle=, pdfauthor=, pdfsubject=, pdfkeywords=}

\newcommand{\ctv}{C_{3v}}

\newcommand{\cfv}{C_{4v}}
\newcommand{\csv}{C_{6v}}
\newcommand{\cnv}{C_{nv}}

\newcommand{\calf}{{\cal F}}

\newcommand{\calc}{{\cal C}}

\newcommand{\gam}[1]{\Gamma_{#1}}

\newcommand{\bra}[1]{\big\langle#1\big|}
\newcommand{\ket}[1]{\big|#1\big\rangle}

\newcommand{\bea}{\begin{eqnarray}}
\newcommand{\enea}{\end{eqnarray}}
\newcommand{\beq}{\begin{equation}}
\newcommand{\eneq}{\end{equation}}

\newcommand{\pdg}[1]{{#1}^{\phantom{\dagger}}}

\newcommand{\W}{{\cal W}}

\newcommand{\bpm}{\begin{pmatrix}}
\newcommand{\epm}{\end{pmatrix}}

\newcommand{\bal}{\begin{align}}
\newcommand{\eal}{\end{align}}

\newcommand{\1}{k^{\sma{(1)}}}

\newcommand{\sma}[1]{\scriptscriptstyle{#1}}

\newcommand{\noc}{n_{\sma{{occ}}}}

\newcommand{\Z}{\mathbb{Z}}

\newcommand{\mo}{\text{-}1}

\newcommand{\qed}{\nobreak \ifvmode \relax \else
      \ifdim\lastskip<1.5em \hskip-\lastskip
      \hskip1.5em plus0em minus0.5em \fi \nobreak
      \vrule height0.75em width0.5em depth0.25em\fi}

\begin{document}

\title{Spin-Orbit-Free Topological Insulators}

\vskip 10pt
\author{
A. Alexandradinata,$^{1}$
B. Andrei Bernevig,$^{1}$ 
}

\address{
$^1$ Department of Physics, Princeton University, Princeton NJ 08544, USA
%\\
%$^2$ Physics Department, Boston University, 
%590 Commonwealth Avenue, Boston, MA 02215, USA
}

\date{\today}

\begin{abstract}
In this lecture for the Nobel symposium, we review previous research on a class of translational-invariant insulators without spin-orbit coupling. These may be realized in intrinsically spinless systems such as photonic crystals and ultra-cold atoms. Some of these insulators have no time-reversal symmetry as well, i.e., the relevant symmetries are purely crystalline. Nevertheless, topological phases exist which are distinguished by their robust surface modes. To describe these phases, we introduce the notions of (a) a \emph{halved}-mirror chirality: an integer invariant which characterizes {half}-mirror planes in the 3D Brillouin zone, and (b) a \emph{bent} Chern number: the traditional Thouless-Kohmoto-Nightingale-Nijs invariant generalized to bent 2D manifolds. Like other well-known topological phases, their band topology is unveiled by the crystalline analog of Berry phases, i.e., parallel transport across certain non-contractible loops in the Brillouin zone. We also identify certain topological phases without any robust surface modes --  they are uniquely distinguished by parallel transport along \emph{bent} loops, whose shapes are determined by the symmetry group. Finally, we describe the Weyl semimetallic phase that intermediates two distinct, gapped phases. 
\end{abstract}

\maketitle

\section{Introduction}

Insulating phases are deemed distinct if they cannot be connected by continuous changes of the Hamiltonian that preserve both the energy gap and the symmetries of the phase; in this sense we say that the symmetry protects the phase. Distinct phases have strikingly different properties -- some nontrivial phases are characterized by robust boundary modes which are particularly accessible through experimental techniques, e.g., STM and ARPES. For many well-known topological insulators (TI's), the existence of boundary bands is in one-to-one correspondence with the topology of the bulk wavefunctions.\cite{fidkowski2011} Examples include the Chern insulator and the quantum spin Hall insulator; these non-interacting insulators fall under ten well-known symmetry classes which are distinguished by time-reversal, particle-hole and chiral symmetries.\cite{schnyder2009} Given the completeness of this classification, attention has shifted to identifying topological phases which rely on other symmetries.\cite{chenfang2012,chenfang2014a,shiozaki} The symmetries which are ubiquitous in condensed matter are the crystal space groups; among them the point groups involve transformations that preserve a spatial point. For example, the reflection symmetry plays an important role in SnTe,\cite{Hsieh2012,Xu2012,tanaka2012} where the mirror Chern number\cite{teo2008} characterizes planes in the 3D Brillouin zone which are invariant under reflection; in short, we call these mirror planes. The Bloch wavefunctions in each mirror plane may be decomposed according to their representations under reflection, and each subspace may exhibit a quantum anomalous Hall effect.\cite{Haldane1988}\\

Including SnTe, all experimentally-realized TI's have thus far been strongly spin-orbit-coupled, resulting in a variety of exotic phenomenon, \emph{e.g.}, Rashba spin-momentum locking on the surface of a TI.\cite{winklerbook} Considerably less attention has been addressed to spin-orbit-free systems, \emph{i.e.}, electronic insulators and semimetals in which spin-orbit coupling is negligibly weak, as well as intrinsically spinless systems such as photonic crystals and ultra-cold atoms. The topological classifications of spin-orbit-free and spin-orbit-coupled systems generically differ. For example, the mirror Chern numbers are trivially zero for any spin-orbit-free system with the same crystalline symmetries as SnTe.\cite{aris2013a} The first proposal of a spin-orbit-free TI with robust surface modes relies on a combination of $n$-fold rotational and time-reversal symmetries, for $n=4$ and $6$.\cite{fu2011} Henceforth, we refer to this combined group as $C_n+T$, where $n \in \{4,6\}$. Topological phases in this symmetry class are sometimes distinguished by their robust surface modes, but not always. A more sensitive probe of their band topology is through the crystalline analog of Berry phases, i.e., parallel transport across certain non-contractible loops in the Brillouin zone. As we elaborate in Sec.\ \ref{sec:c4t}, the topological phases without surface modes are uniquely distinguished by parallel transport along \emph{bent} loops, whose shapes are determined by the symmetry group.\cite{aris2014a} Besides parallel transport,\cite{aa2012} the entanglement spectrum is also known to distinguish topological phases without boundary modes, in various symmetry classes.\cite{turner2010A,hughes2011,turner2012,chenfang2013a}\\

\begin{figure}[H]
\centering
\includegraphics[width=10 cm]{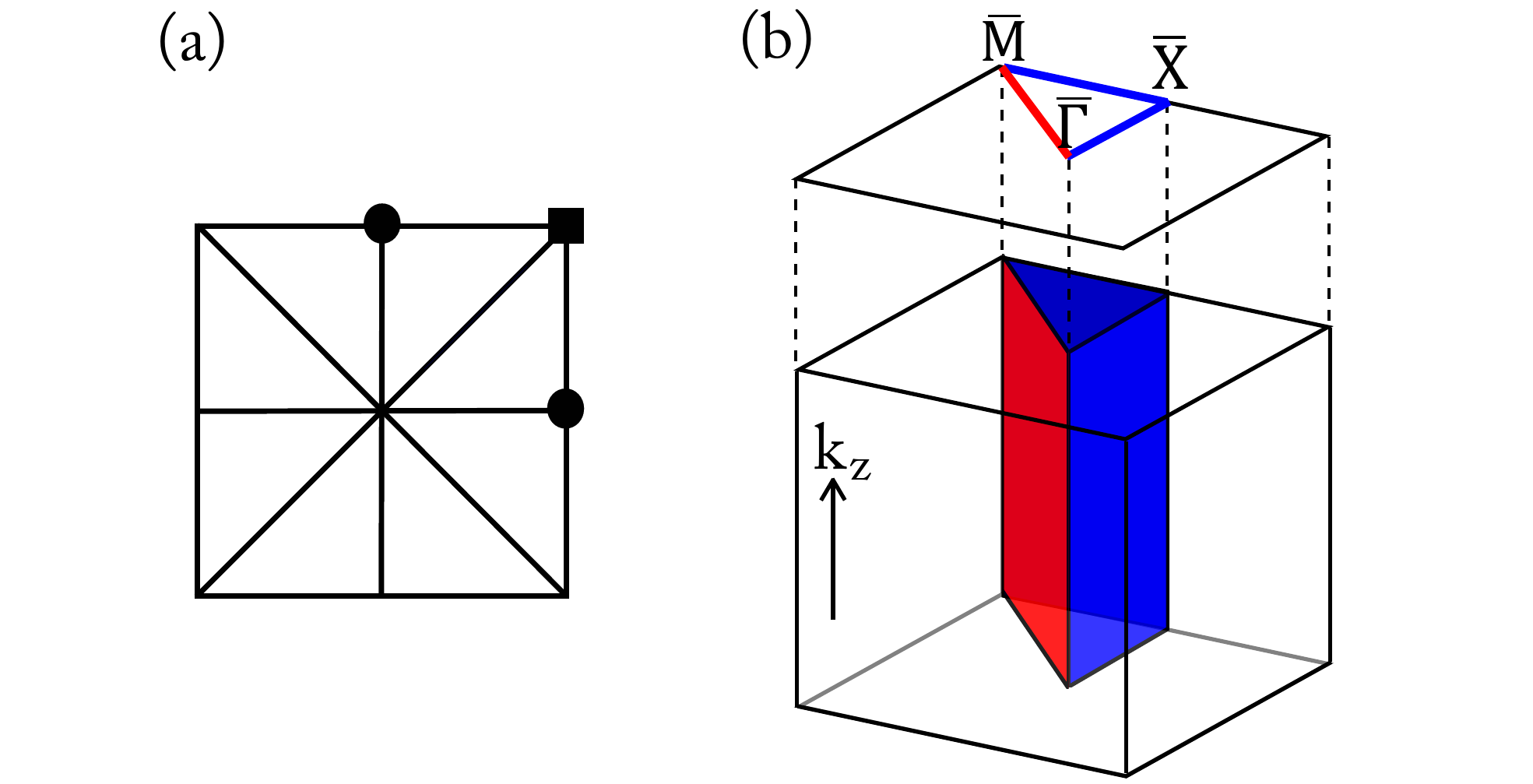}
\caption{  (a) Top-down view of tetragonal BZ with $\cfv$ symmetry; our line of sight is parallel to the rotational axis. Reflection-invariant planes are indicated by solid lines. Except the line through $\Gamma$, all non-equivalent $C_n$-invariant lines are indicated by circles for $n=2$, and squares for $n=4$. Bottom of (b): Half-mirror planes (HMP's) in the 3D Brillouin zone (BZ) of a tetragonal lattice with $\cfv$ symmetry. Red: HMP$_1$. Blue: HMP$_2$. We define a bent Chern number on the triangular pipe with its ends identified. Top of (b): Non-black lines are half-mirror lines (HML's) in the corresponding 2D BZ of the 001 surface; each HML connects two distinct $C_4$-invariant points.}\label{HMP}
\end{figure}

As with many other well-known TI's,\cite{fu2006,roy2009} time-reversal symmetry (TRS) plays an essential role in protecting the $C_n+T$ phase. Can one do away with TRS entirely? Indeed, it is now known that spin-orbit-free topological phases can exist without TRS; the relevant symmetries are \emph{purely} crystalline. These phases include the first TI protected by symmorphic point groups,\cite{aris2013a} as well as the first TI protected by non-symmorphic space groups.\cite{chaoxingliu2013}. In this short review, we establish sufficient criteria for a symmetry group to protect robust boundary modes. By applying this criteria to the 32 crystallographic point groups, we exhaustively identify the $\cnv$ groups, for $n =3,4$ and $6$, as being able to support robust surface modes without additional symmetries. For example, the symmetries of $\cfv$ are described in Fig.\ \ref{HMP}(a). To describe these $\cnv$ systems, we introduce the notions of (a) a \emph{halved}-mirror chirality: an integer invariant which characterizes {half}-mirror planes in the 3D Brillouin zone (BZ), and (b) a \emph{bent} Chern number: the traditional TKNN invariant\cite{thouless1982} generalized to bent 2D manifolds (illustrated in Fig.\ \ref{HMP}(b)).  We find that a Weyl semimetallic phase intermediates two gapped phases with distinct halved chiralities. In Sec.\ \ref{sec:c4v} of this review, we characterize the $\cfv$ systems, leaving a more comprehensive treatment of the other $\cnv$ groups to Ref.\ \onlinecite{aris2013a}.\\

\section{$\cfv$ insulators and semimetals} \label{sec:c4v}

\begin{figure}[H]
\centering
\includegraphics[width=10 cm]{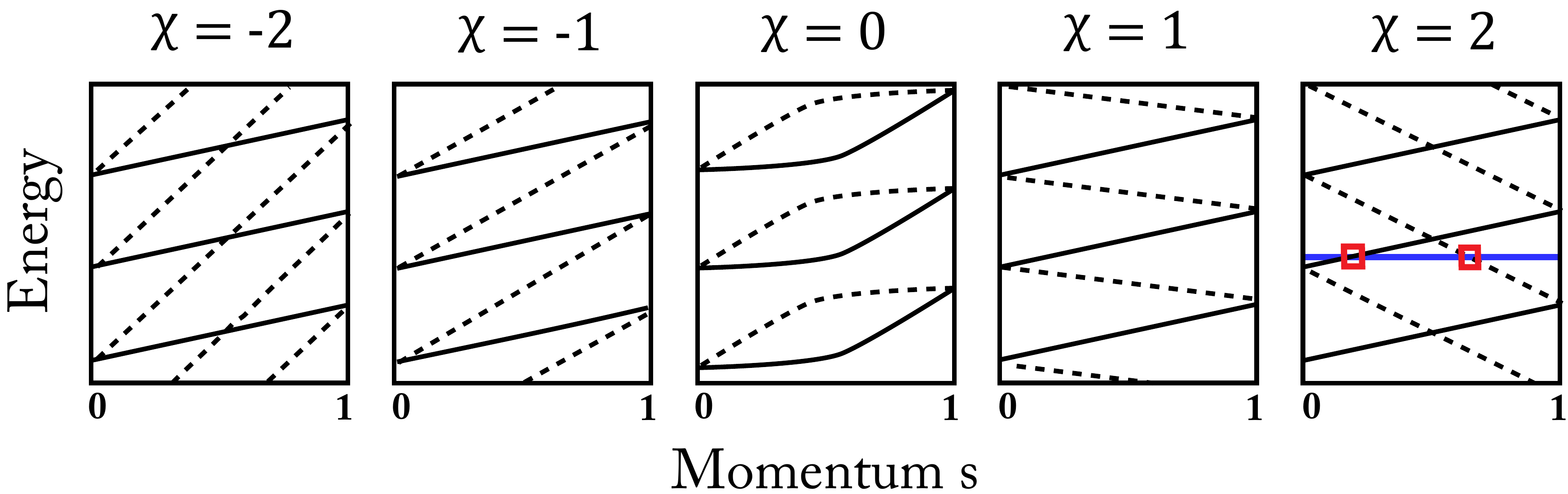}
\caption{ Distinct connectivities of the (001) surface bands along the half-mirror-lines. Black solid (dotted) lines indicate surface bands with eigenvalue $+1$ ($-1$) under reflection $M_i$; crossings between solid and dotted lines are robust due to reflection symmetry. For simplicity, we have depicted all degeneracies at momenta $s=0$ and $s=1$ as dispersing linearly with momentum. For $\cfv$, such crossings are in reality quadratic.\cite{aris2013a}}\label{connectivity}
\end{figure}

Given that the mirror Chern number vanishes in spin-orbit-free systems with $\cfv$ symmetry, how else can surface modes be made robust? Progress is made by identifying two $C_4$-invariant points ($\bar{\Gamma}$ and $\bar{M}$) in the (001) surface BZ; each $C_4$-invariant point is mapped to itself under a four-fold rotation, up to translations by a reciprocal lattice vector, as we illustrate in Fig.\ \ref{HMP}(a). In addition, the black lines that intersect $\bar{\Gamma}$ (resp.\ $\bar{M}$) in Fig.\ \ref{HMP}(a) are invariant under reflection, indicating that there are additional symmetries at this point, e.g., the reflection $M_1$ maps $(x,y) \rightarrow (y,x)$. The combined symmetries of $\bar{\Gamma}$ (resp.\ $\bar{M}$) form the little group\cite{tinkhambook} $\cfv$. At these points, the symmetry enforces that $(p_x,p_y)$ and $(d_{xz},d_{yz})$ orbitals are doubly-degenerate. That is, each pair of orbitals transforms in the two-dimensional irreducible representation (irrep) of $\cfv$. Indeed, if we choose a basis ($p_x\pm p_y$) that diagonalizes $M_1$ with eigenvalue $\pm 1$, clearly $C_4$ transforms $p_x +p_y \rightarrow -(p_x-p_y)$ and vice versa. We then propose that surface bands of $\cfv$ systems assume topologically distinct structures on the red line which connects $\bar{\Gamma}$ and $\bar{M}$, as illustrated in Fig.\ \ref{HMP}(b). We call the red line a half-mirror line (HML$_1$) because it does not close into a circle (modulo reciprocal lattice vectors), yet all Bloch wavefunctions on the red line may be diagonalized by a \emph{single} reflection operator ($M_1$). Suppose we parametrize HML$_1$ with $s_1 \in [0,1]$, where $s_1=0$ (resp.\ $1$) at $\bar{\Gamma}$ (resp.\ $\bar{M}$). At $s_1=0$ and $1$, we have just established that (001) surface bands (with $(p_x,p_y)$ character) form doubly-degenerate pairs with opposite mirror eigenvalues, irrespective of whether the system has TRS. For representations without spin, we may label bands with mirror eigenvalue $+1$ $(-1)$ as mirror-even (mirror-odd). Given these constraints at $s_1=0$ and $1$, there are $\Z$ ways to connect mirror-even bands to mirror-odd bands, as illustrated schematically in Fig.\ \ref{connectivity}; it should be noticed that band intersections do not split as they correspond to bands with opposite mirror eigenvalues. We define the halved mirror chirality $\pdg{\chi}_1 \in \Z$ as the \emph{difference} in number of mirror-even chiral modes  with mirror-odd chiral modes; if $\pdg{\chi}_1 \neq 0$, the surface bands robustly interpolate across the energy gap. $\pdg{\chi}_1$ may be easily extracted by inspection of the surface energy-momentum dispersion: first draw a constant-energy line within the bulk energy gap and parallel to the HML, {e.g.}, the blue line in Fig.\ \ref{connectivity}. At each intersection with a surface band, we calculate the sign of the group velocity $dE/ds_1$, and multiply it with the eigenvalue under reflection $M_1$. Finally, we sum this quantity over all intersections along HML$_1$ to obtain $\pdg{\chi}_1$. In Fig.\ \ref{connectivity}, we find two intersections as indicated by red squares, and $\pdg{\chi}_1 = (1)(1) + (-1)(-1)=2$. The $\Z$-classification of (001) surface bands relies on two-dimensional irreps in the surface BZ; on surfaces which break $\cfv$ symmetry, the surface bands transform in one-dimensional irreps, and cannot assume topologically distinct structures. \\

In addition to HML$_1$, the blue line in Fig.\ \ref{HMP}(b) is a second half-mirror line (HML$_2$), and states along HML$_2$ are characterized by an independent invariant $\pdg{\chi}_2$. HML$_2$ is the union of two mirror lines, $\bar{\Gamma}-\bar{X}$ and $\bar{X}-\bar{M}$, which are related by a $\pi/2$ rotation. States along $\bar{\Gamma}-\bar{X}$ are invariant under the reflection $M_y: y \rightarrow -y$, while along $\bar{X}-\bar{M}$ the relevant reflection is $M_x: x \rightarrow -x$. The product of these orthogonal reflections is a $\pi$ rotation ($C_2$) about $\hat{z}$, thus $M_x = C_2\,M_y$. Since both $(p_x,p_y)$ orbitals are odd under a $\pi$-rotation, an eigenstate of $M_x$ with eigenvalue $\eta$ also diagonalizes $M_y$ with eigenvalue ${-\eta}$, e.g., $p_x$ is even under $M_y$ but odd under $M_x$. In effect, each Bloch wavefunction along HML$_2$ can be labelled by the quantum number of a \emph{single} reflection operator $M_y \equiv M_2$. The second invariant $\pdg{\chi}_2$ is then the {difference} in number of mirror-even ($M_2=+1$) chiral modes with mirror-odd ($M_2=-1$) chiral modes, in a straightforward generalization of $\pdg{\chi}_1$.\\

Thus far we have described the halved chirality $\pdg{\chi}_i$ as a topological property of surface bands along HML$_i$, but we have not addressed how $\pdg{\chi}_i$ is encoded in the \emph{bulk} wavefunctions. Taking $\hat{z}$ to lie along the rotational axis, each HML$_i$ in the surface BZ is the $\hat{z}$-projection of a half-mirror-plane (HMP$_i$) in the 3D BZ, as illustrated in Fig.\ \ref{HMP}(b). Each HMP connects two \emph{distinct} $C_4$-invariant lines, and all Bloch wavefunctions in a HMP may be diagonalized by a single reflection operator. HMP$_i$ is parametrized by $t_i \in [0,1]$ and $k_z \in (-\pi,\pi]$, where $t_i=0$ $(1)$ along the first (second) $C_4$-invariant line. Then the halved mirror chirality has the following gauge-invariant expression by bulk wavefunctions:\cite{aris2013a}
\bal \label{mirrorchirality}
\pdg{\chi}_i = \frac{1}{2\pi}\,\int_{\text{HMP}_i} dt_i\,dk_z\,(\,\calf_{e} - \calf_{o}\,) \in \Z.
\end{align}
$\calf_{e}$ ($\calf_{o}$) is defined as the Berry curvature\cite{berry1984,zak1982,zak1989} of occupied bands as contributed by the mirror-even (-odd) subspace. \\

\begin{figure}[H]
\centering
\includegraphics[width=10 cm]{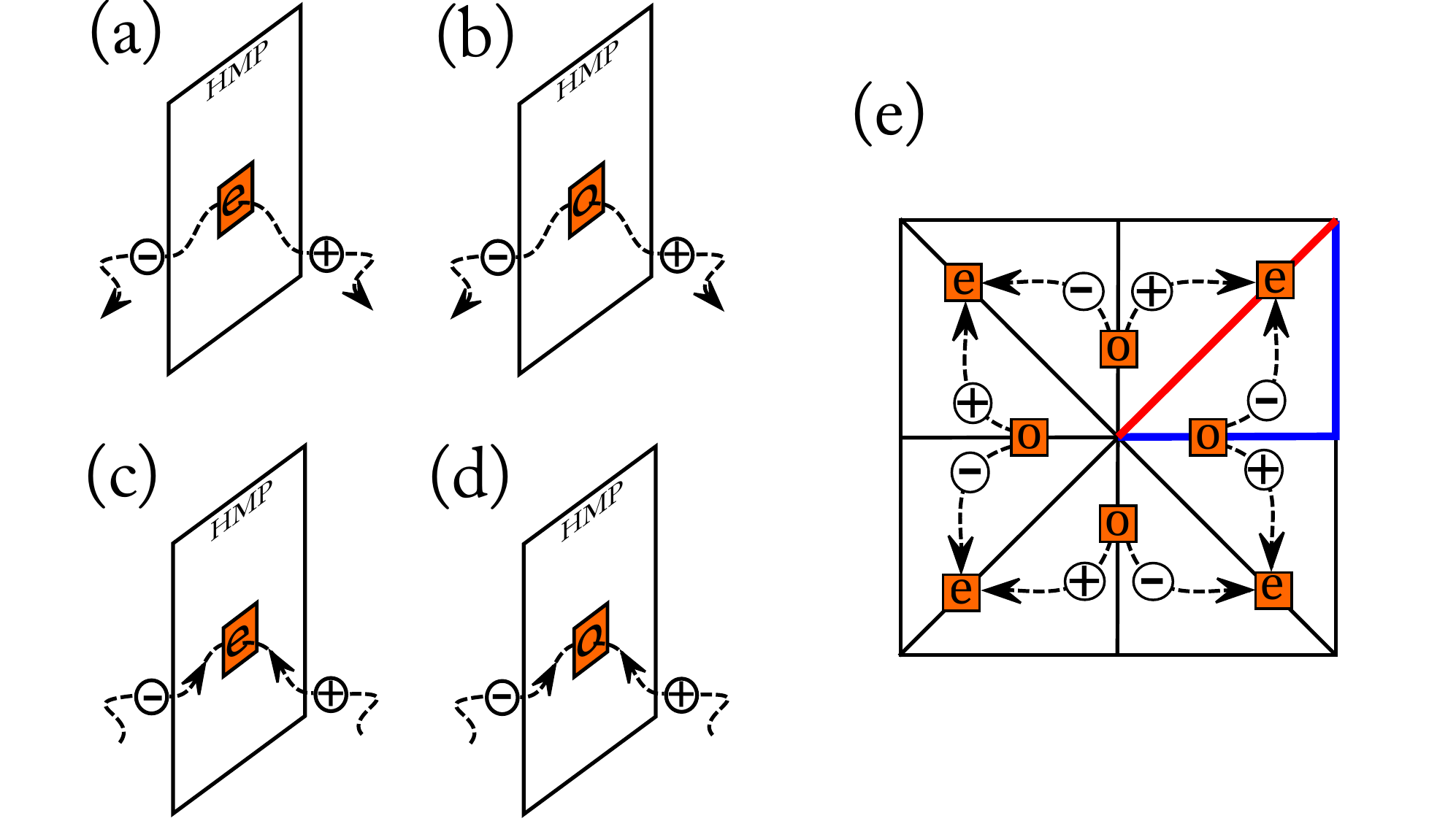}
\caption{(a) to (d) illustrate how the halved chirality of a HMP may change. The direction of arrows indicate whether Weyl nodes are created or annihilated. + (-) labels a Berry monopole with positive (negative) charge; $e$ ($o$) labels a crossing in the HMP between mirror-even (-odd) bands. (e) We provide a top-down view of the trajectories of Weyl nodes, in the transition between two distinct gapped phases. Our line of sight is parallel to $\hat{z}$.  Black lines indicate mirror faces, while colored lines specially indicate HMP's, with the same color legend as in Fig.\ \ref{HMP}(b).}\label{weyltrajectories}
\end{figure}

How do transitions occur between gapped phases with different $\{\pdg{\chi}_i\}$? The invariants $\{\pdg{\chi}_i\}$ are well-defined so long as bulk states in the HMP's are gapped, which is true of $\cfv$ insulators. These invariants may also be used to characterize $\cfv$ semimetals, so long as the gaps close away from the HMP's. When gaps close, the band touchings are generically Weyl nodes,\cite{Murakami2007B,wan2010} though exceptions exist with a conjunction of time-reversal and inversion symmetries.\cite{burkov2011} The chirality of each Weyl node is its Berry charge, which is positive (negative) if the node is a source (sink) of Berry flux. By the Nielsen-Ninomiya theorem, the net chirality of all Weyl nodes in the BZ is zero.\cite{NIELSEN1981} To make progress, we divide the BZ into `unit cells', such that the properties of one `unit cell' determine all others by symmetry. As seen in Fig.\ \ref{HMP}(b), each `unit cell' resembles the interior of a triangular pipe; it is known as the orbifold $T^3/\cfv$. The net chirality of an orbifold can be nonzero, and is determined by the Chern number on the 2D boundary of the orbifold. As the boundary resembles the surface of a triangular pipe, we call the invariant a bent Chern number ($\calc_{12}$). $\cfv$ systems are thus described by three invariants $(\pdg{\chi}_{1}, \pdg{\chi}_{2}, \calc_{12})$, which are related by $\text{parity}[ \,\pdg{\chi}_{1} + \pdg{\chi}_{2}\,] = \text{parity}[\, \calc_{12}\,].$ If $\sum_{i=1}^2 \pdg{\chi}_{i}$ is odd, $\calc_{12}$ must be nonzero due to an odd number of Weyl nodes within the orbifold, which implies the system is gapless. If the system is gapped, $\sum_{i=1}^2 \pdg{\chi}_{i}$ must be even. However, the converse is not implied. This parity constraint may be understood in light of a Weyl semimetallic phase that intermediates two gapped phases with distinct halved chiralities. There are four types of events that alter the halved chirality $\pdg{\chi}_i$ of HMP$_i$; we explain how Weyl nodes naturally emerge in the process. (i) Suppose the gap closes between two mirror-even bands in HMP$_i$. Around this band touching, bands disperse linearly within the mirror plane, and quadratically in the direction orthogonal to the plane. Within HMP$_i$, the linearized Hamiltonian around the band crossing describes a massless Dirac fermion in the even representation of reflection. If the mass of the fermion inverts sign, $\int_{HMP_i} \calf_{e}/2\pi$ changes by $\eta \in \{\pm 1\}$, implying that $\pdg{\chi}_i$ also changes by $\eta$ through Eq.\ (\ref{mirrorchirality}). This quantized addition of Berry flux is explained by a splitting of the band-touching into two Weyl nodes of opposite chirality, and on opposite sides of the mirror face (Fig.\ \ref{weyltrajectories}(a)). In analogy with magnetostatics, the initial band touching describes the nucleation of a dipole, which eventually splits into two opposite-charge monopoles; the flux through a plane separating two monopoles is unity. (ii) The same argument applies to the splitting of dipoles in the mirror-odd subspace, which alters $\int_{\sma{HMP_i}} \calf_{o}/2\pi$ by $\kappa \in \{\pm 1\}$, and  $\pdg{\chi}_i$ by $-\kappa$. For (iii) and (iv), consider two opposite-charge monopoles which converge on HMP$_i$ and annihilate, causing $\pdg{\chi}_i$ to change by unity. The sign of this change is determined by whether the annihilation occurs in the mirror-even or odd subspace. The transition between two distinct gapped phases is then characterized by a transfer of Berry charge between two distinct HMP's (Fig.\ \ref{weyltrajectories}(e)). In the intermediate semimetallic phase, the experimental implications include Fermi arcs on the (001) surface.\cite{wan2010,halasz2012,chen2012}

\section{$C_n+T$ insulators and semimetals} \label{sec:c4t}

\begin{figure}[H]
\centering
\includegraphics[width=12 cm]{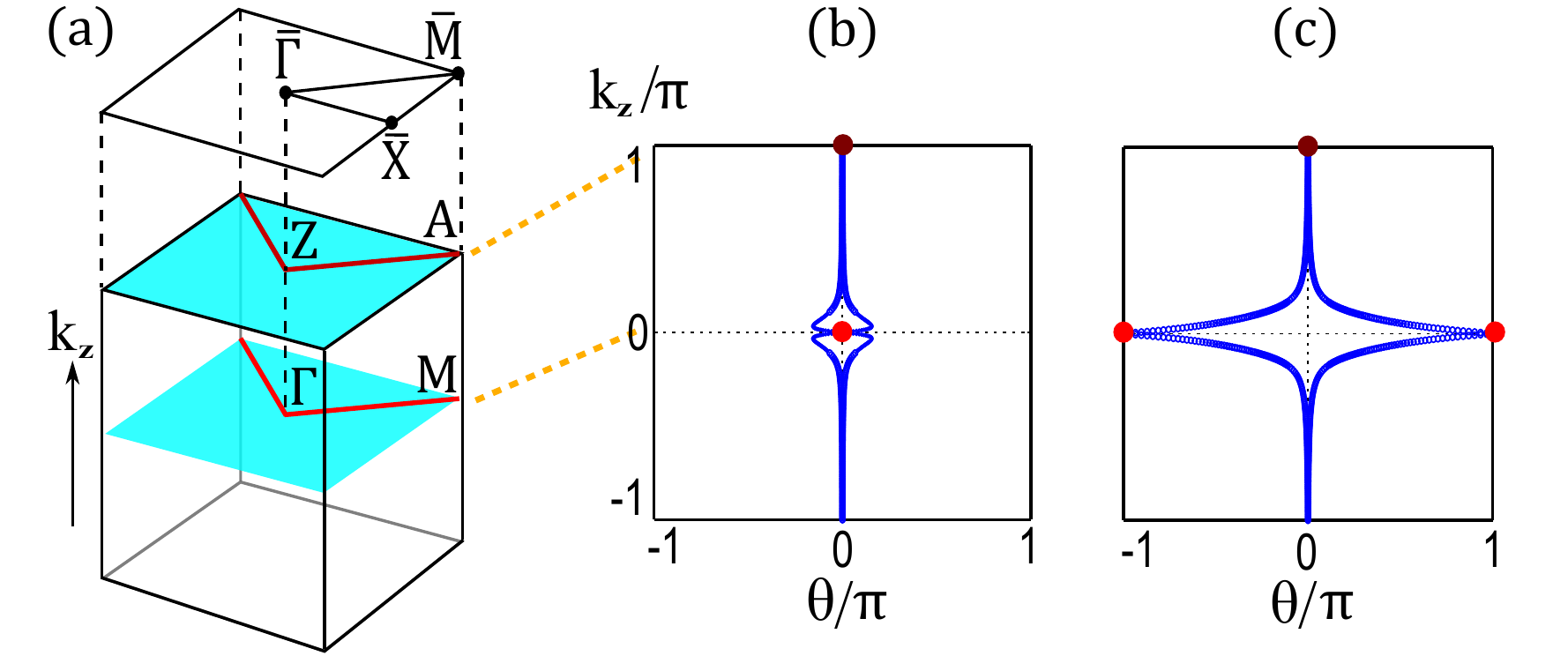}
\caption{Bottom of (a) illustrates certain bent loops in the 3D Brillouin zone of a simple tetragonal lattice. Top of (a): (001)-surface BZ of the same lattice. (b) and (c): Berry-phase spectrum of trivial and strong phases respectively, for the model of Eq.\ (\ref{cfvmodel}).}\label{fig:BZsWloops}
\end{figure}

The topological connectivity of surface bands in $\cfv$ systems relied on a criterion which may be generalized: all we need are two high-symmetry points on the boundary BZ, which admit multi-dimensional irreps at each point. Suppose we relaxed the reflection symmetries in the group $\cfv$, and now added time-reversal symmetry. The resultant group ($C_4+T$) also has two-dimensional irreps at $C_4$-invariant momenta $\bar{\Gamma}$ and $\bar{M}$. For example, the $p_x\pm ip_y$ orbitals are eigenstates of four-fold rotation, while time-reversal maps $p_x+ip_y \rightarrow p_x-ip_y$. This two-fold degeneracy in a spin-orbit-free system is analogous to the Kramer's degeneracy in spin-orbit coupled systems. Similarly, there exists two distinct gapped phases which are distinguished by robust surface modes;\cite{fu2011} we refer to one as the trivial phase and the other a strong topological phase. Our case study of the $C_4+T$ insulator is motivated by two issues: (i) It is unclear if these two phases are physically distinguishable if we experimentally probe the bulk instead of the surface. (ii) In this review, we highlight the existence of a third topological phase which does not manifest robust surface modes. Does this `weak phase' have any physical consequence? One answer to (i) and (ii) may be found through holonomy, i.e., parallel transport along certain non-contractible loops in the BZ.\cite{leone2011,ekert2000,recati2002}  An electron transported around a loop acquires a Berry-Zak phase,\cite{berry1984,zak1982,zak1989} which has recently been measured by interference in cold atom experiments.\cite{atala2013} For the purpose of unveiling the bulk topology of all three phases, we find that not all non-contractible loops work. Straight loops are commonly studied in the geometric theory of polarization, due to their relation with Wannier functions;\cite{kingsmith1993,Taherinejad2013} however, these loops cannot identify our weak phase. Instead, we propose that all three phases are distinguished by parallel transport along \emph{bent} loops, whose shapes are determined by the symmetry group -- they are illustrated in Fig. \ref{fig:BZsWloops}(a) for the $C_4+T$ group, while the $C_6+T$ case is discussed in Ref.\ \onlinecite{aris2014a}. \\

\begin{figure}[H]
\centering
\includegraphics[width=12 cm]{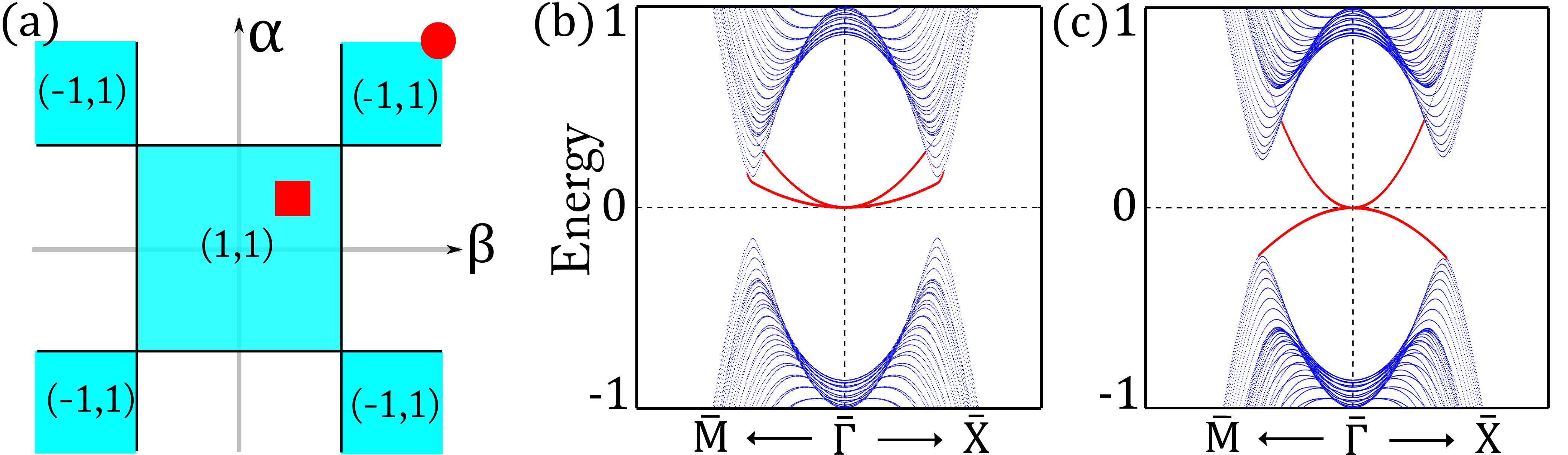}
\caption{(a) Phase diagram of $C_4+T$ model\ (\ref{cfvmodel}), as a function of parameters $\alpha$ and $\beta$. Blue (uncolored) regions correspond to gapped (gapless Weyl) phases.\cite{aris2013a} The weak indices in each gapped phase are indicated by $(\Gamma_4(0),\Gamma_4(\pi))$. The blue square in the center is approximately bound by $|\alpha|<2$ and $|\beta|<2$. The 001-surface spectrum is plotted for two representative points on the phase diagram: (b) for $\alpha=\beta=1$, and (c) for $\alpha=\beta=4$. $\bar{\Gamma}$,$\bar{M}$ and $\bar{X}$ are high-symmetry momenta defined in Fig.\ \ref{fig:BZsWloops}(a).}\label{fig:c4vmodel}
\end{figure}

It is known from Ref.\ \onlinecite{fu2011} that the $C_4+T$ insulator is characterized by two $\Z_2$ indices $\{\Gamma_4(\bar{k}_z)\}$, for $\bar{k}_z \in \{0,\pi\}$; $\Gamma_4(0) \in \{+1,-1\}$ describes the Bloch wavefunctions in the plane $k_z=0$. In analogy with the spin-orbit-coupled $\Z_2$ insulator,\cite{fu2011,fu2007b,roy2009a} $\{\Gamma_4(0),\Gamma_4(\pi)\}$ shall be referred to as weak indices. $\Gamma_4(0)=\Gamma_4(\pi)=1$ (resp. ${-1}$) corresponds to the trivial (resp. weak) phase, and $\Gamma_4(0)={-\Gamma_4(\pi)}$ describes a strong phase. The product $\Gamma_4(0)\,\Gamma_4(\pi)$ is a strong index that determines the absence or presence of robust surface modes on the 001 surface; we take $\hat{z}$ to lie along the principal $C_4$ axis. We give these weak indices a physically transparent interpretation from the perspective of holonomy. For illustration, we consider a model on a tetragonal lattice that is composed of two interpenetrating cubic sublattices.  Our tight-binding basis comprises of $(p_x,p_y)$ orbitals, which transform in the two-dimensional irrep of $C_4+T$. The Bloch Hamiltonian is
\bal \label{cfvmodel}
 H(\boldsymbol{k}) =  \big[\mo + 8f_1(\boldsymbol{k})\big]\,\Gamma_{03}  + 2 f_2(\boldsymbol{k})\,\gam{11} + \alpha f_4(\boldsymbol{k})\,\gam{01}  +  \beta f_5(\boldsymbol{k})\,\gam{32} + 2 f_6(\boldsymbol{k})\,\gam{12},
\end{align}
where $f_1 = 3-\text{cos}(k_x)-\text{cos}(k_y)-\text{cos}(k_z)$, $f_2 = 2-\text{cos}(k_x)-\text{cos}(k_y)$, $f_3=\text{cos}(k_z)$, $f_4 = \text{cos}(k_y)-\text{cos}(k_x)$,  $f_5 = \text{sin}(k_x)\,\text{sin}(k_y)$ and $f_6 = \text{sin}(k_z)$. In $\Gamma_{ab} = \sigma_a \otimes\tau_b$, $\sigma_i$ and $\tau_i$ are Pauli matrices for $i \in \{1,2,3\}$, while $\sigma_0$ and $\tau_0$ are identities in each 2D subspace.  $\ket{\sigma_3= \pm 1,\tau_3=+1}$  label $\{p_x \pm i p_y\}$ orbitals on one sublattice, and $\ket{\sigma_3= \pm 1,\tau_3=-1}$  label $\{p_x \mp i p_y\}$ orbitals on the other. This Hamiltonian is four-fold symmetric: $\Gamma_{33}\,H(k_x,k_y,k_z)\,\Gamma_{33} = H({-k_y},k_x,k_z\,)$, and time-reversal symmetric: $\Gamma_{10}\,H(\boldsymbol{k})^*\,\Gamma_{10} = H(-\boldsymbol{k})$. The ground state of (\ref{cfvmodel}) comprises its two lowest-lying bands. The phase diagram of this model is plotted in Fig.\ \ref{fig:c4vmodel}(a) for different parametrizations of Eq.\ (\ref{cfvmodel}). \\

To probe the bulk topology, we perform parallel transport along a bent loop that connects two $C_4$-invariant points. We define $l_4(0)$ as the loop connecting $M-\Gamma-M$ in the $k_z=0$ plane, and $l_4(\pi)$ connects $A-Z-A$ in the $k_z=\pi$ plane. They are respectively depicted by red and brown lines in Fig.\ \ref{fig:BZsWloops}(a). Let us denote the eigenstates of the Bloch Hamiltonian by $H(\boldsymbol{k})\,\ket{u_{i,\boldsymbol{k}}} = \varepsilon_{i,\boldsymbol{k}}\,\ket{u_{i,\boldsymbol{k}}}$, where $i$ is a band index. The matrix representation of holonomy is known as the Wilson loop, and it is the path-ordered exponential of the Berry-Wilczek-Zee connection $A(\boldsymbol{k})_{ij} = \bra{u_{i,\boldsymbol{k}}}\,\nabla_{\boldsymbol{k}}\,\ket{u_{j,\boldsymbol{k}}}$: $\W[l] = \text{exp}\,\big[{-\int_l dl \cdot A(\boldsymbol{k})}\,\big].$\cite{wilczek1984,berry1984} Here, $l$ denotes a loop and $A$ is a matrix with dimension equal to the number ($\noc$) of occupied bands. The gauge-invariant spectrum of $\W[l]$ is also known as the Berry-phase factors ($\{\text{exp}(i \vartheta)\}$).\cite{zak1982,zak1989,aa2012} We show that the spectrum of $\W[l_4(\bar{k}_z)]$ encodes the weak index $\Gamma_4(\bar{k}_z)$. If we define $d_{4}$ is the number of $-1$ eigenvalues in the spectrum of $\W[l_{4}]$, we find that this number is always even. Furthermore, the weak indices $\{\Gamma_4(0),\Gamma_4(\pi)\}$ are related to $\{d_{4}(0),d_4(\pi)\}$ by
\bal \label{2dz2inv}
\Gamma_4(\bar{k}_z) = i^{d_4(\bar{k}_z)} \in \{1,\mo\}; \;\; \;\bar{k}_z \in \{0,\pi\}.
\end{align}
This weak index is equivalent to an alternative formulation in Ref.\ \onlinecite{fu2011}, where it is expressed as an invariant involving the Pfaffian of a matrix.\cite{aris2014a} Our formulation through holonomy reveals a geometric connection with the theory of matrices in $SO(\noc)$: as we elaborate in Ref.\ \onlinecite{aris2014a}, the parity of $d_4(\bar{k}_z)/2$ specifies one of two classes of a special rotation, which is in one-to-one correspondence with two sectors of ground states in the $k_z=\bar{k}_z$ plane. Furthermore, we are able to derive the spectrum of $\W[l_4(\bar{k}_z)]$ for any number of occupied bands.\cite{aris2014a} In a minimal model with $\noc=2$, there are only two possibilities ($\W[l_4(\bar{k}_z)]=\pm I$), which are distinguished by the parity of $d_{4}(\bar{k}_z)/2$. \\

Let us demonstrate how to realize both classes of $\W[l_4(\bar{k}_z)]$ in the model (\ref{cfvmodel}). We consider a family of loops $\{l_4(k_z)\}$ in planes of constant $k_z$, such that $l_4(0)$ is the red line in Fig.\ \ref{fig:BZsWloops}(a), and all other loops project to $l_4(0)$ in $\hat{z}$. In the trivial phase (parametrized by $\alpha=\beta=1$), we find $\W[l_4(0)]=\W[l_4(\pi)]=I$, or equivalently $\Gamma_4(0)=\Gamma_4(\pi)=+1$. The eigenvalues of $\W[l_4(k_z)]$ interpolate between $\{1,1\}$ (at $k_z=0$) to $\{1,1\}$ (at $k_z=\pi$), as illustrated in Fig.\ \ref{fig:BZsWloops}(b). The absence of surface modes on the line $\bar{M}-\bar{\Gamma}-\bar{X}$ is demonstrated in Fig.\ \ref{fig:c4vmodel}(b). In comparison, the strong phase ($\alpha=\beta=4$) is characterized by $\W[l_4(0)]=-\W[l_4(\pi)]=-I$, or $\Gamma_4(0)={-\Gamma_4(\pi)}={-1}$; its surface modes are illustrated in Fig.\ \ref{fig:c4vmodel}(c). As $k_z$ is varied from $0$ to $\pi$ in Fig.\ \ref{fig:BZsWloops}(c), the Berry phases $\{\vartheta(k_z)\}$ interpolate across the maximal range $(-\pi,\pi]$. We say that the Berry phases exhibit spectral flow if they interpolate across their maximally-allowed range as we tune a BZ parameter. Spectral flow is a unifying trait shared by many TI's, including the Chern insulator,\cite{Haldane1988,thouless1982,qi2006,ProdanJMP2009} the quantum spin Hall insulator,\cite{yu2011,soluyanov2011} and the inversion-symmetric TI.\cite{aa2012} In comparison, the weak phase is distinguished by $\W[l_4(0)]=\W[l_4(\pi)]={-I}$, which clearly does not exhibit spectral flow. Finally, we point out that a different choice of Wilson loop can identify the strong index $\Gamma_4(0)\Gamma_4(\pi)$,\cite{Taherinejad2013} but cannot individually distinguish the weak indices: $\Gamma_4(0)$ and $\Gamma_4(\pi)$.

\section{Conclusion}

We propose sufficient criteria for surface modes whose robustness rely on a symmetry group. Minimally, (i) the symmetry must be unbroken by the presence of the surface. Additionally, either (ii) a reflection symmetry exists so that mirror-subspaces can display a quantum anomalous Hall effect, or (iii) there exist at least two high-symmetry points in the surface BZ, which admit higher-than-one dimensional irreps of the symmetry group. In addition to predicting new topological materials, these criteria are also satisfied by the well-known SnTe class,\cite{Hsieh2012} and also the $\Z_2$ insulators.\cite{fu2006,kane2005A,kane2005B,bernevig2006a,bernevig2006c,koenig2007,fu2007b,moore2007,roy2009,roy2009a,hsieh2008} Among the $32$ crystallographic point groups, only the $C_n$ and $\cnv$ groups are preserved for a surface that is orthogonal to the rotational axis.\cite{tinkhambook} Though all $\cnv$ groups satisfy (ii), the lack of spin-orbit coupling implies only $\ctv$ systems can have nonvanishing mirror Chern numbers.\cite{aris2014a} While all $C_n$ groups by themselves only have one-dimensional irreps, the $C_4$ or $C_6$ group satisfies (iii) in combination with TRS, as is known for the topological crystalline insulators introduced in Ref.\ \onlinecite{fu2011}. Finally, only a subset of the $\cnv$ groups possess two-dimensional irreps which satisfy (iii): $\ctv, \cfv$ and $\csv$.\cite{aris2013a}\\

\noindent\textbf{Acknowledgments}\\

Chen Fang and Matthew Gilbert played an essential role in formulating the theory of $\cnv$ insulators, whose short review we have presented here. AA and BAB were supported by Packard Foundation, 339-4063--Keck Foundation, NSF CAREER DMR-095242, ONR - N00014\text{-}11\text{-}1-0635 and  NSF-MRSEC DMR-0819860 at Princeton University. This work was also supported by DARPA under SPAWAR Grant No.: N66001\text{-}11\text{-}1-4110. CF is supported by ONR-N0014-11-1-0728 for salary and DARPA-N66001-11-1-4110 for travel. MJG acknowledges support from the ONR under grant N0014-11-1-0728 and N0014-14-1-0123, a fellowship from the Center for Advanced Study (CAS) at the University of Illinois.

\end{document}